# Radiative Correction to the Casimir Energy for Massive Scalar Field in The Network

M. A. Valuyan*
*Department of Physics, Se. C., Islamic Azad University, Semnan, Iran*

In this paper, we compute the leading-order and first-order radiative corrections to the Casimir energy of a massive Lorentz-violating scalar field governed by a $\phi^4$ interaction on a network. For simplicity, the network is chosen to consist of three edges connected at a single central junction, with the scalar field defined in $1+1$ dimensions on each edge. Dirichlet boundary conditions are imposed at the outer ends of the edges, thereby confining the field on the network. Beyond addressing the massive case of a Lorentz-violating scalar field, a key novelty of this work lies in the calculation of the radiative corrections to the Casimir energy using position-dependent counterterms. These counterterms emerge from a systematic renormalization program that consistently incorporates the effects of boundary conditions into the renormalization procedure. To eliminate divergences arising from vacuum energy contributions, we employ the box subtraction scheme in conjunction with cutoff regularization. Our results show that both the leading-order and first-order Casimir energies are negative, regardless of the presence or absence of Lorentz-violating effects, and are in agreement with general physical expectations.

## I. INTRODUCTION

The Casimir effect is a striking manifestation of quantum vacuum fluctuations, arising from modifications of the zero-point energy induced by boundary conditions, background fields, or nontrivial geometries. First predicted by H. B. G. Casimir in 1948 for parallel perfectly conducting plates [1], the Casimir effect has since become a cornerstone of quantum field theory (QFT), with broad applications ranging from condensed matter physics [2] and nanotechnology [3, 4] to biophysics [5–7], cosmology [8–12], and high-energy physics [13]. Owing to its strong dependence on boundary conditions, topology [14], geometry, and dimensionality [15], the Casimir energy provides a sensitive theoretical probe of the interplay between geometry and quantum fluctuations [16]. Following the pioneering work of Bordag *et al.* on radiative corrections to the Casimir energy, extensive studies have been carried out in this direction [17–20]. Radiative corrections have been investigated for various quantum fields under different boundary conditions and geometrical settings [21], as well as in curved manifolds and cosmological backgrounds [22, 23]. A central issue in these analyses concerns the appropriate choice of counterterms within the renormalization program [24]. In many early studies, the form of the counterterms was assumed to be independent of the boundary conditions imposed on the quantum fields, and free-space counterterms were employed universally [25]. Subsequent works attempted to incorporate boundary effects into the renormalization procedure, for instance by introducing additional surface counterterms localized at the boundaries [26, 27]. However, such approaches did not fully capture the influence of nontrivial boundary conditions. A more systematic framework was developed by Gousheh *et al.* [28], in which all boundary effects are consistently incorporated into the renormalization program. Within this scheme, a perturbative expansion is performed and the renormalization conditions naturally lead to the emergence of position-dependent counterterms. As is well known, counterterms are introduced to renormalize the bare parameters of the Lagrangian, and position-dependent counterterms fulfill this role as effectively as their free-space counterparts. Their spatial dependence arises through the Green's function of the system, which itself encodes the effects of geometry and boundary conditions. Consequently, all boundary-induced modifications are automatically incorporated into the renormalization procedure. From this perspective, when boundary conditions are imposed on quantum fields, the renormalization program—and in particular the structure of the counterterms—must be influenced accordingly. The indiscriminate use of free counterterms, regardless of the imposed boundary conditions, may therefore be inadequate and can even lead to unphysical divergent results [29].

In recent years, increasing attention has been devoted to quantum fields defined on networks, graphs, and branched structures [30–33]. Such systems, commonly referred to as quantum graphs or networks, provide effective models for a wide variety of physical systems, including quantum wires, photonic crystals, branched nanotubes, mesoscopic devices, molecular structures, and polymer networks. In contrast to simple geometries, networks possess junctions, branching points, and nontrivial connectivity, which significantly modify the spectral properties of the underlying differential operators. As a result, both classical and quantum vacuum energies in these systems exhibit novel features. The Casimir energy of a massless scalar field on a network with Dirichlet and Neumann boundary conditions has been previously investigated in Ref. [34]. Motivated by these developments, in the present work we extend previous studies by computing both the leading-order and the first-order radiative correction to the Casimir energy for a massive, self-interacting scalar field described by $\phi^4$ theory on a simple network with three edges. We further generalize our analysis to the case of Lorentz-violating scalar fields. While radiative corrections to the Casimir energy have been extensively studied for massive and massless scalar and electromagnetic fields in flat geometries, curved manifolds, and spaces with defects, their investigation in network-like configurations remains relatively unexplored. The computation of radiative corrections in such networks therefore constitutes one of the main novel aspects of this work. Another important feature of our approach is the systematic use of position-dependent counterterms in the renormalization of the theory. Our method is based on constructing the Green's function of the network in a manner fully consistent with the boundary conditions at the edges and the junction. Analytic continuation and appropriate regularization schemes are then employed to extract finite, physically meaningful contributions to the vacuum energy. We also analyze how Lorentz symmetry breaking modifies the dispersion relations and Green's functions, and how these modifications affect both the leading-order Casimir energy and its radiative correction. The paper is organized as follows: we first introduce the network model and derive the Green's function on each edge. We then compute the leading-order Casimir energy for a massive scalar field, followed by the first-order radiative correction arising from the $\phi^4$ interaction. The effects of Lorentz symmetry breaking on both contributions are subsequently examined. Finally, we summarize our results and discuss possible extensions or ideas, including more general networks and alternative boundary conditions.

## II. DESCRIPTION OF THE MODEL

In this section, we outline the general framework for computing the Casimir energy in a simple network. To this end, we begin with the Lagrangian density of the $\phi^4$ theory in $1+1$ dimensions,

$$\mathcal{L} = \frac{1}{2}[\partial_\mu\phi(x)\partial^\mu\phi(x) + \alpha(\mathbf{u}.\partial\phi(x))^2] - \frac{1}{2}m_0^2\phi^2(x) - \frac{\lambda_0}{4!}\phi(x)^4, \tag{1}$$

where $x = (t, x_j)$, and the dimensionless parameter $\alpha \ll 1$ characterizes the strength of Lorentz-symmetry breaking. The parameter $m_0$ denotes the bare mass of the quantum field, while $\lambda_0$ is the bare coupling constant. As illustrated in Fig. (1), the coordinate $x_{j=\{1,2,3\}}$ represents the distance from the central node $N$ along each edge $E_1$, $E_2$, and $E_3$ of the network, respectively. The vector $\mathbf{u}$ specifies the direction of Lorentz-symmetry violation. Since the scalar field is considered in $1+1$ dimensions, two distinct choices for $\mathbf{u}$ are possible. Selecting $\mathbf{u} = (1, 0)$ corresponds to time-like Lorentz-symmetry breaking, whereas choosing $\mathbf{u} = (0, 1)$ leads to space-like Lorentz violation. We assume that the massive scalar field described by the Lagrangian in Eq. (1) is defined on the three edges of the network. Under this assumption, the equation of motion derived from the Lagrangian in Eq. (1) on each edge of the network is given by

$$[\Box + \alpha(\mathbf{u}.\partial)^2 + m_0^2]\phi(x) = 0. \tag{2}$$

where $\Box = \partial_t^2 - \partial_{x_j}^2$. As a first step, and for the sake of simplicity, we restrict our analysis to the Lorentz-symmetric scalar field case ($\alpha = 0$). The effects of Lorentz violation on the Casimir energy, for both time-like and space-like symmetry breaking, will be discussed subsequently. Solving the equation of motion given in Eq. (2) and imposing Dirichlet boundary conditions at the endpoints $x_j = L_j$ lead to the following expression for the wave function:

$$\phi(x_j, t) = \phi_k(x_j)e^{-i\omega t}, \tag{3}$$

where $\phi_k(x_j) = A_j \sin[k(x_j - L_j)]$, and $\omega_k^2 = k^2 + m_0^2$, with $\omega_k$ denoting the wave number. The general form of the junction conditions at the node $N$ for networks with $\mathcal{N}$ edges was previously introduced in Ref. [34] and is given by

$$\phi_i|_N = \phi_j|_N, \qquad \sum_{j=1}^{\mathcal{N}} \partial_{x_j}\phi_j|_N = 0. \tag{4}$$

Using the wave-function expression given in Eq. (3), the junction conditions for a network with three edges, as illustrated in Fig. (1), reduce to the following set of equations:

$$A_1 \sin kL_1 = A_2 \sin kL_2 = A_3 \sin kL_3, \qquad A_1 \cos kL_1 + A_2 \cos kL_2 + A_3 \cos kL_3 = 0. \tag{5}$$

Combining the two equations arising from the junction conditions given in Eq. (5) leads to the following constraint equation, which determines the spectrum of the wave vector $k$:

$$\Delta(k) = \cot kL_1 + \cot kL_2 + \cot kL_3 = 0. \tag{6}$$

The constraint function $\Delta(k)$ possesses a discrete set of poles located at $k = \{\frac{n\pi}{L_1}, \frac{n\pi}{L_2}, \frac{n\pi}{L_3}\}$ with $n \in \mathbb{Z}^+$. These poles correspond to the quantized spectral modes determined by the characteristic length scales $L_1$, $L_2$, and $L_3$ of the network. To renormalize the bare parameters $m_0$ and $\lambda_0$ appearing in the Lagrangian given in Eq. (1), we rescale the field according to $\phi = \sqrt{Z}\,\phi_r$, which leads to

$$\mathcal{L} = \frac{1}{2}[(\partial_\mu\phi_r)^2 + \alpha(\mathbf{u}.\partial\phi_r)^2] - \frac{1}{2}m^2\phi_r^2 - \frac{\lambda}{4!}\phi_r^4 + \frac{1}{2}[(\partial_\mu\phi_r)^2 + \alpha(\mathbf{u}.\partial\phi_r)^2]\delta_z - \frac{1}{2}\delta_m\phi_r^2 - \frac{\delta_\lambda}{4!}\phi_r^4, \tag{7}$$

where $Z = 1 + \delta_Z$ is the field-strength renormalization factor. Furthermore, the quantities $\delta_m = Zm_0^2 - m^2$ and $\delta_\lambda = Z^2\lambda_0 - \lambda$ denote the mass and coupling-constant counterterms, respectively. For the above Lagrangian, the Feynman rules associated with these counterterms are given by

$$\begin{aligned} -\!\otimes\!- &= i\big[(p^2 - \alpha(\mathbf{u}.p)^2)\delta_Z - \delta_m\big], \\ \times\!\!\!\!\otimes &= -i\delta_\lambda. \end{aligned} \tag{8}$$

In the framework of renormalized perturbation theory, as expressed in Eq. (7), the first few terms of the perturbative expansion of the two-point function can be represented symbolically as

$$\text{[diagram]}_{x_1\,x_2} = \text{[diagram]}_{x_1\,x_2} + \text{[diagram]}_{x_1\,x\,x_2} + \text{[diagram]}_{x_1\,x\,x_2}, \tag{9}$$

where $\text{[diagram]}_{x_1\,x\,x_2}$ denotes the counterterm. To determine the values of the counterterms, we impose the following renormalization conditions:

$$\begin{aligned} -\!\bullet\!- &= \frac{i}{p^2 - \alpha(\mathbf{u}.p)^2 - m^2} + \text{(the terms regular at } p^2 - \alpha(\mathbf{u}.p)^2 = m^2), \\ \times\!\!\!\!\bullet &= -i\lambda \qquad\qquad (\text{at } s = 4m^2, t = u = 0). \end{aligned} \tag{10}$$

where the parameters $s$, $t$, and $u$ specify the type of scattering channel. As is well known, the channel is determined from the topology of the corresponding Feynman diagram, and each channel yields a characteristic angular dependence of the cross section. Using Eq. (9) together with the renormalization conditions in Eq. (10), and retaining terms up to first order in the coupling constant $\lambda$, the general expression for the mass counterterm is obtained as

$$\delta_m(x) = \frac{-i}{2}\,\text{[diagram]}_{x_1\,x\,x_2} = \frac{-\lambda}{2}G(x;x), \tag{11}$$

where $G(x;x)$ denotes the Green's function and $x = (x_j, t)$. The details of the computation of $G(x;x)$ are presented in Appendix A. Up to first order in the coupling constant $\lambda$, the renormalization conditions yield vanishing counterterms, i.e., $\delta_\lambda = \delta_Z = 0$. In the next step, we obtain an expression for the total vacuum energy of the network up to first order in $\lambda$. Hence, we have

$$\begin{aligned} E^{(1)}_{\text{vac.}} &= i\sum_{j=1}^{3}\int_0^{L_j} dx_j\left(\frac{1}{8}\,\text{[diagram]} + \frac{1}{2}\,\text{[diagram]} + ...\right) \\ &= i\sum_{j=1}^{3}\int_0^{L_j} dx_j\left(\frac{-i\lambda}{8}G^2(x;x) - \frac{-i}{2}\delta_m(x)G(x;x)\right), \end{aligned} \tag{12}$$

where $L_j$ denotes the length of each edge of the network. The superscript (1) on the vacuum energy indicates the first order in the coupling constant $\lambda$. Combining Eqs. (11) and (12) for the network shown on the left-hand side of Fig. (1) yields

$$E^{(1)}_{\text{vac.}} = \frac{-\lambda}{8}\sum_{j=1}^{3}\int_0^{L_j} dx_j G^2(x;x). \tag{13}$$

As is well known, in the conventional definition of the Casimir energy, the vacuum energies of two configurations—with and without the boundary conditions—are subtracted. In the present work, we employ a modified subtraction scheme, which is a slight modification of the box subtraction scheme originally introduced by T. H. Boyer [22, 35]. In this scheme, a network similar to the original one is constructed, and the vacuum energies of the two networks are subtracted (see Fig. 1). Consequently, the Casimir energy can be expressed as

$$E_{\text{Cas.}} = E^{(0)}_{\text{Cas.}} + E^{(1)}_{\text{Cas.}} = \lim_{L'_j\to\infty}\sum_{j=1}^{3}\left\{\left[E^{(0)}_{\text{vac.}}(L_j) - E^{(0)}_{\text{vac.}}(L'_j)\right] + \left[E^{(1)}_{\text{vac.}}(L_j) - E^{(1)}_{\text{vac.}}(L'_j)\right]\right\} \tag{14}$$

where $E^{(0)}_{\text{vac}}(L_j)$ and $E^{(0)}_{\text{vac}}(L'_j)$ denote the zeroth-order (zero-point) vacuum energies of the left and right networks shown in Fig. 1, respectively. Likewise, $E^{(1)}_{\text{vac}}(L_j)$ and $E^{(1)}_{\text{vac}}(L'_j)$ represent the corresponding first-order vacuum energies of the left and right configurations. The difference between these first-order contributions yields the radiative correction to the Casimir energy. The lengths of the edges in the second network (i.e., the right network in Fig. (1)) are ultimately taken to be infinite. Imposing this limit ensures that the second network plays the role of Minkowski space; hence, the subtraction of the vacuum energies of the two networks reproduces the standard definition of the Casimir energy. Introducing a second, similar network for the computation of the Casimir energy adds extra parameters (namely $L'_j$), which serve as additional regulators. These regulators provide more degrees of freedom for regularizing the divergences that arise during the calculation, making the cutoff regularization technique applicable. Moreover, the presence of these extra parameters clarifies the removal of divergences by allowing a more controlled adjustment of the cutoff values. In the next section, we present the detailed computation of the leading-order Casimir energy using this regularization scheme. Subsequently, the radiative correction to the Casimir energy is presented in Section IV.

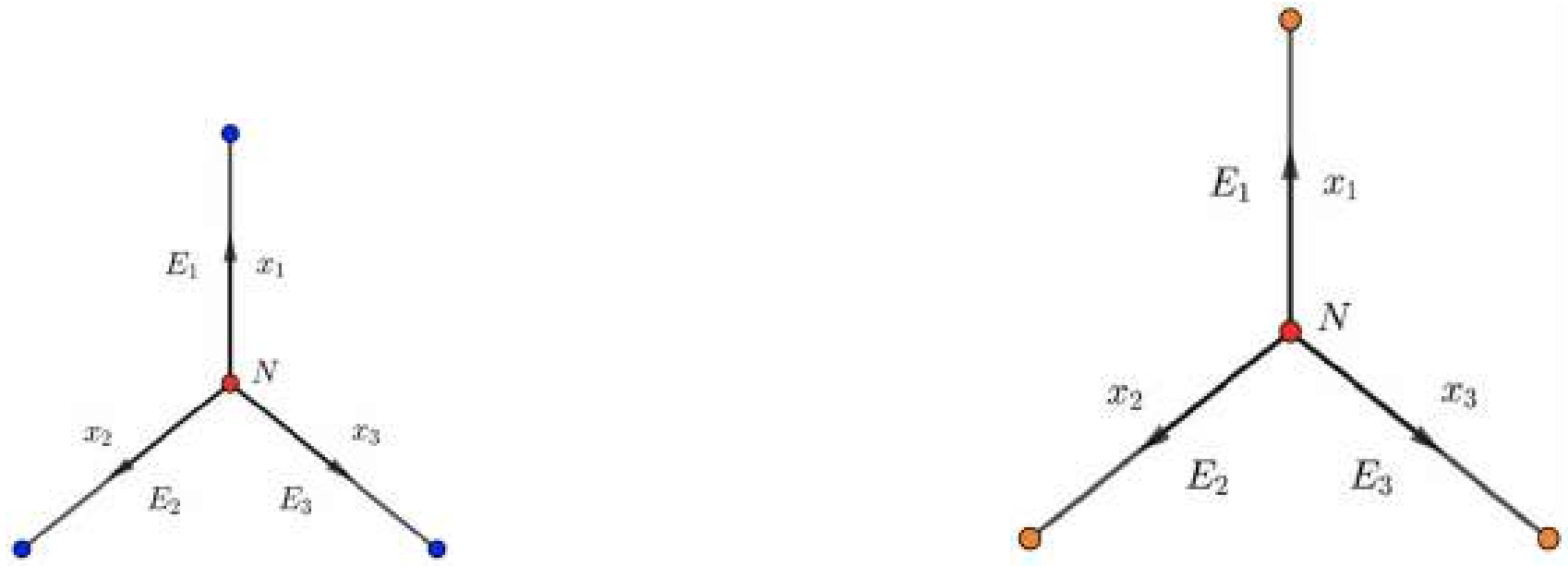

FIG. 1. Schematic representation of the simplest network consisting of three edges joined at a single node. The junction condition given in Eq. (4) is imposed at the central node (red point), while Dirichlet boundary conditions are applied at the endpoints of the edges (blue points in the left panel and orange points in the right panel). The coordinate $x_i$ along edge $E_i$ is measured from the central node $N$, such that $x_i = 0$ at $N$. The length of edge $E_i$ is $L_i$ for $i = \{1, 2, 3\}$.

## III. LEADING-ORDER CASIMIR ENERGY

Starting from the standard expression for the zero-point energy and applying Eq. (14), we obtain

$$E^{(0)}_{\text{Cas.}} = \lim_{L'_j\to\infty}\Delta E^{(0)}_{\text{vac.}} = \lim_{L'_j\to\infty}\left\{\frac{1}{2}\sum_{k}\sqrt{k^2+m^2} - \frac{1}{2}\sum_{k'}\sqrt{k'^2+m^2}\right\}. \tag{15}$$

Here, the summation over $k$ denotes a sum over all roots of the constraint function $\Delta(k) = 0$, which correspond to the allowed modes of the original network (left panel of Fig. 1). Similarly, the summation over $k'$ in Eq. (15) runs over all roots of $\Delta(k') = 0$ associated with the auxiliary network (right panel of Fig. (1)). As noted in Eq. (15), the lengths of the edges in the auxiliary network are ultimately taken to infinity, so that this network plays the role of Minkowski space. Since the spectrum of roots does not exhibit a simple or regular structure, direct summation is not feasible. In such cases, contour integration provides a powerful analytical method, and the application of Cauchy's theorem offers an efficient and rigorous approach to evaluate the sum. Therefore, we convert the discrete sums over $k$ and $k'$ into contour integrals. Using Cauchy's theorem, the sum over the allowed wave numbers $\omega_k = \sqrt{k^2+m^2}$, where $k$ are the roots of $\Delta(k) = 0$, can be written as

$$\frac{1}{2}\sum_{k}\sqrt{k^2+m^2} = \frac{1}{2\pi i}\oint\frac{\sqrt{z^2+m^2}}{2}d\ln\Delta(z) + \frac{1}{2}\sum_{n=1}^{\infty}\sum_{j=1}^{3}\sqrt{P^2_{n,j}+m^2}, \tag{16}$$

where $P_{n,j}=\frac{n\pi}{L_j}$ denotes the poles of the constraint function $\Delta(k)$. To perform the summation over all roots of $\Delta(k)$ in the complex $z$-plane, we enclose the roots by a contour shaped as a semicircle in the right half-plane. This contour consists of a line segment along the imaginary axis and a curved arc of radius $R\to\infty$, which together enclose all real roots of $\Delta(k)$ located on the positive real axis. Using this contour integration, the first term on the right-hand side of Eq. (16) separates into two contributions: one arising from the integration along the imaginary axis (the line segment of the semicircle) and the other originating from the integration along the curved arc as $R\to\infty$. Consequently, the first term on the right-hand side of Eq. (16) becomes

$$\frac{1}{2\pi i}\oint g(z)dz=\frac{-1}{\pi}\int_m^\infty g(iy)dy+\frac{R}{2\pi}\int_{\frac{-\pi}{2}}^{\frac{\pi}{2}} g(Re^{i\theta})e^{i\theta}d\theta, \tag{17}$$

where $g(z)=\frac{\sqrt{z^2+m^2}}{2}\frac{\Delta'(z)}{\Delta(z)}$. The contribution of the second term in Eq. (17), which corresponds to the curved part of the contour, vanishes in the limit $R\to\infty$. Consequently, only the contribution from the line segment along the imaginary axis (the first integral on the right-hand side of Eq. (17)) remains. Based on Eq. (15), we apply the same procedure to the auxiliary network. The only distinction is that, upon taking the limit $L'_j\to\infty$, the contribution arising from the line segment of the semicircle vanishes. Taking this fact into account and using Eqs. (16) and (17), the subtraction of vacuum energies in Eq. (15) can be simplified to

$$\Delta E^{(0)}_{\text{vac.}}=\frac{-1}{\pi}\int_m^\infty g(iy)dy+\left[\frac{1}{2}\sum_{n=1}^{\infty}\sum_{j=1}^{3}\sqrt{P_{n,j}^2+m^2}-\{L_j\to L'_j\}\right]. \tag{18}$$

As mentioned above, the first term inside the brackets on the right-hand side of Eq. (18) corresponds to the sum over all poles of the constraint function $\Delta(k)$. This summation is divergent and therefore requires regularization. Unlike the sum over the roots of $\Delta(k)$, the present summation runs over its regular poles. To regularize the divergences arising from these poles, we employ the following form of the Abel–Plana summation formula (APSF):

$$\sum_{n=1}^{\infty}\mathcal{F}(n)=\frac{-1}{2}\mathcal{F}(0)+\int_0^\infty\mathcal{F}(x)dx+i\int_0^\infty\frac{\mathcal{F}(it)-\mathcal{F}(-it)}{e^{2\pi t}-1}dt, \tag{19}$$

where the first and second terms on the right-hand side are referred to as the "*zero term*" and the "*integral term*", respectively. Typically, the divergent contributions of the summation are contained in these two terms. The last term on the right-hand side of Eq. (19) is commonly known as the branch-cut term and usually provides the convergent contribution to the summation. Applying the APSF in Eq. (19) converts the bracketed expression in Eq. (18) to

$$\frac{1}{2}\sum_{n=1}^{\infty}\sum_{j=1}^{3}\sqrt{P_{n,j}^2+m^2}-\{L_j\to L'_j\}=\sum_{j=1}^{3}\left[\frac{-m}{4}+\frac{L_j}{2\pi}\int_0^\infty\sqrt{\xi^2+m^2}d\xi+B(L_j)\right]-\{L_j\to L'_j\}, \tag{20}$$

where $B(L_j)$ denotes the branch-cut term of the APSF. The term $\frac{-m}{4}$ in the bracket of Eq. (20) cancels exactly with its counterpart originating from the auxiliary network. The second term in the bracket of Eq. (20) corresponds to the integral term of the APSF and is divergent. However, using the cutoff regularization scheme, its contribution is canceled by the analogous term from the auxiliary network. To implement this cancellation, we introduce a cutoff $\Lambda$ for the upper limit of the integral term in Eq. (20). Similarly, we introduce a cutoff $\Lambda'$ for the corresponding integral term of the auxiliary network. Evaluating the integrals and choosing the cutoffs such that $\Lambda\sqrt{\Lambda^2+m^2}+m^2\ln\left(\frac{\Lambda+\sqrt{\Lambda^2+m^2}}{m}\right)=\frac{L'_j}{L_j}\left[\Lambda'\sqrt{\Lambda'^2+m^2}+m^2\ln\left(\frac{\Lambda'+\sqrt{\Lambda'^2+m^2}}{m}\right)\right]$ ensures that the integral terms in Eq. (20) cancel exactly. Consequently, the only remaining contribution from the APSF is the branch-cut term, which is

$$B(L_j)=\frac{L_j}{\pi}\int_m^\infty\frac{\sqrt{\tau^2-m^2}}{e^{2L_j\tau}-1}d\tau=\sum_{n=1}^{\infty}\frac{mK_1(2nL_jm)}{2\pi n}, \tag{21}$$

where $K_1(x)$ is the modified Bessel function of the second kind. As indicated in Eq. (15), taking the limit $L'_j\to\infty$ is the final step of the calculation. In this limit, the contribution of the branch-cut term $B(L'_j)$ vanishes. Consequently, the final expression for the Casimir energy can be written as

$$E^{(0)}_{\text{Cas.}}=\frac{-1}{\pi}\int_m^\infty g(iy)dy+\sum_{j=1}^{3}\sum_{n=1}^{\infty}\frac{mK_1(2nL_jm)}{2n\pi}. \tag{22}$$

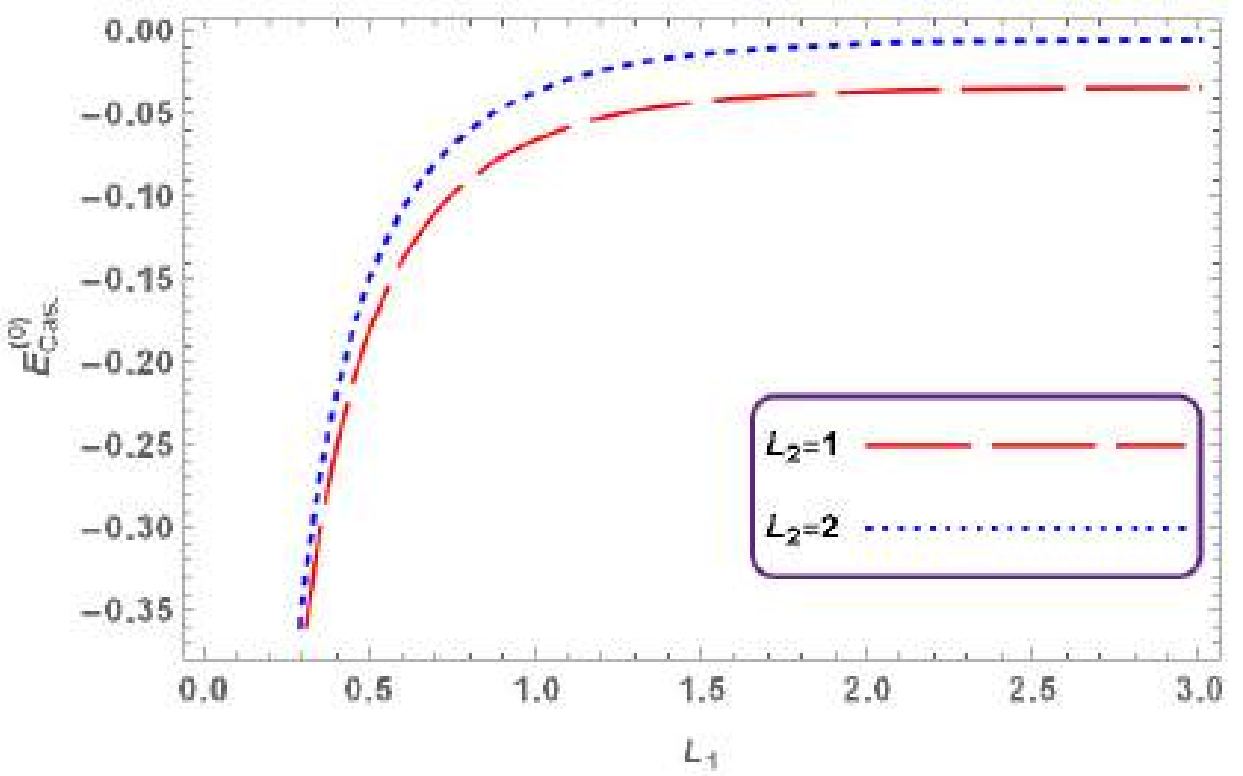

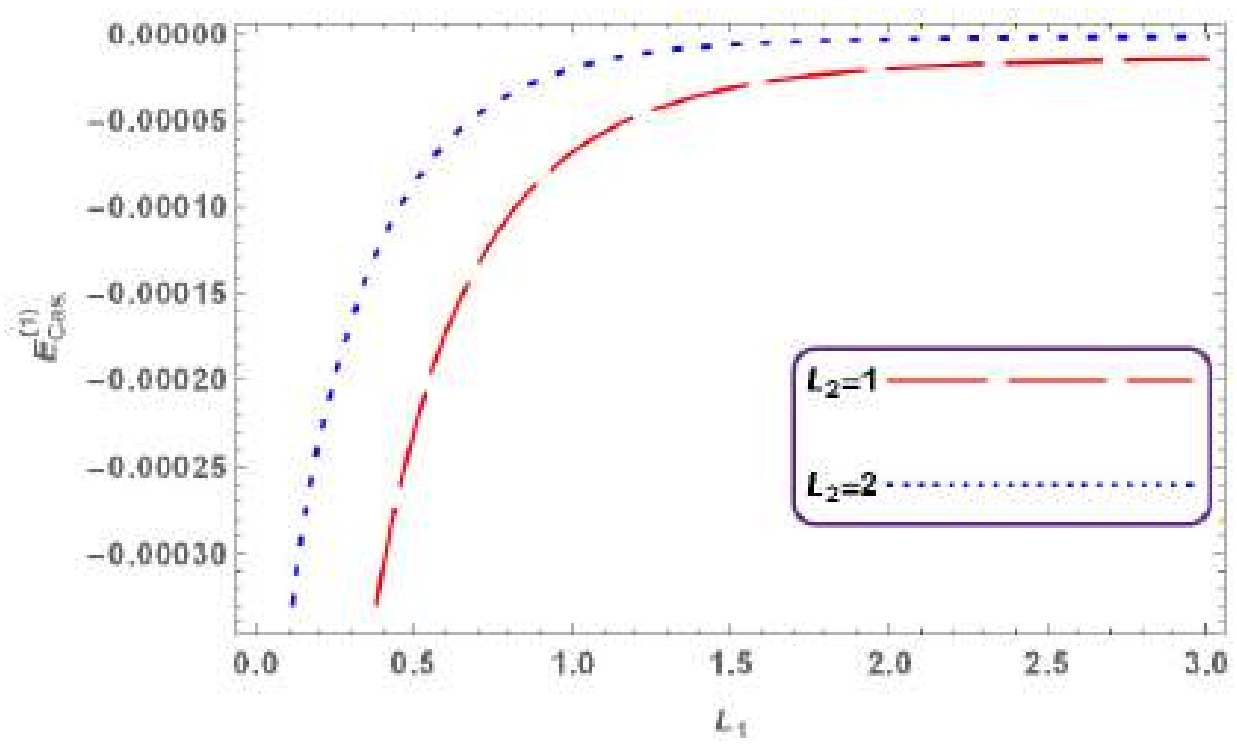

FIG. 2. Left panel: Leading-order Casimir energy for a massive scalar field, given by Eq. (22), plotted as a function of the edge length $L_1$. The remaining edge lengths are fixed at $L_2 = \{1, 2\}$ and $L_3 = 2$. Right panel: First-order radiative correction to the Casimir energy for a massive scalar field, given by Eq. (29), plotted as a function of $L_1$, with $L_2 = \{1, 2\}$ and $L_3 = 2$. In all plots, the field mass is $m = 1$ and the coupling constant is $\lambda = 0.1$.

When the Lorentz-symmetry breaking parameter is switched on ($\alpha \neq 0$), two distinct cases arise. Choosing $\mathbf{u} = (1, 0)$ corresponds to time-like Lorentz violation. Repeating the above computation for this case shows that the frequency is modified as $\omega \to \tilde{\omega} = \frac{\omega}{\sqrt{1+\alpha}}$. Consequently, time-like Lorentz violation affects the leading-order Casimir energy only through an overall multiplicative factor of $\frac{1}{\sqrt{1+\alpha}}$. Choosing $\mathbf{u} = (0, 1)$ corresponds to space-like Lorentz violation. In this case, the same computation yields a Casimir energy expression that coincides with the Lorentz-symmetric result ($\alpha = 0$), except that the edge lengths are rescaled as $L_j \to \tilde{L}_j = \frac{L_j}{\sqrt{1-\alpha}}$. Two plots of the Casimir energy as a function of one of the edge lengths of the network shown in the left panel of Fig. (1) are presented. The effects of time-like and space-like Lorentz violation, relative to the Lorentz-symmetric case, are illustrated in the left panel of Fig. (4). Although Lorentz violation does not change the sign of the Casimir energy, it produces a significant quantitative deviation from the Lorentz-symmetric value.

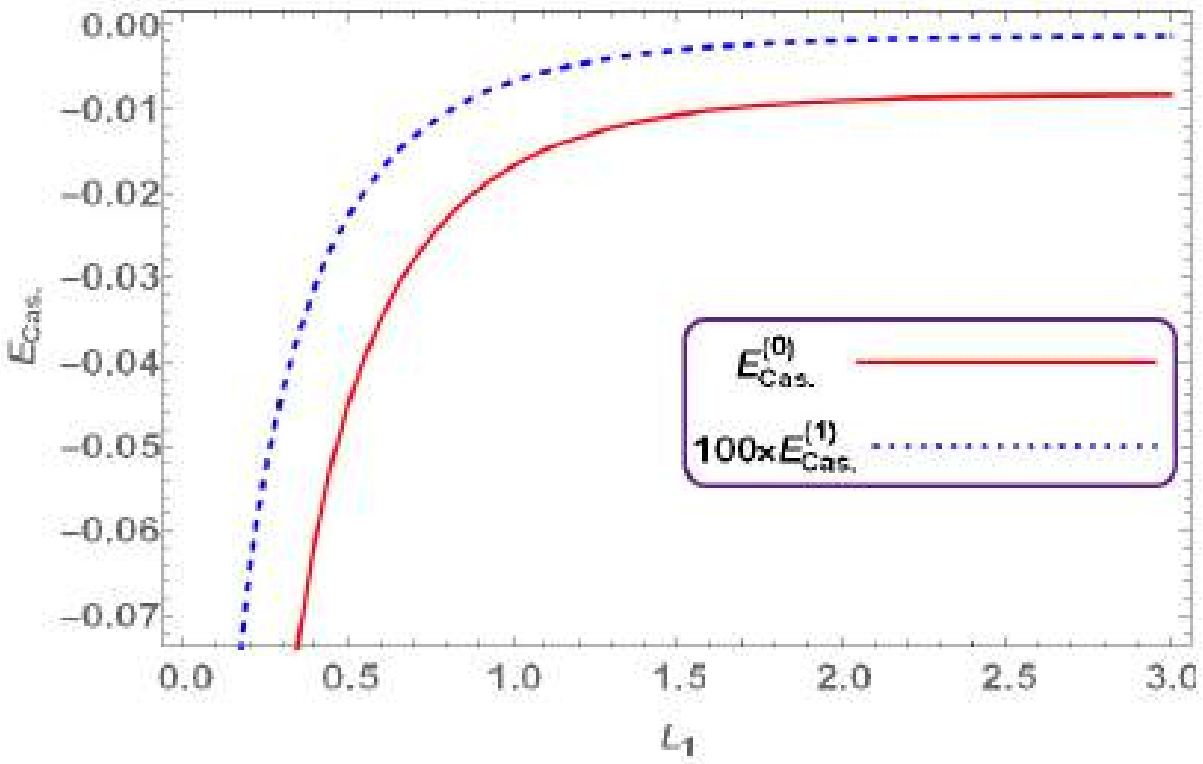

FIG. 3. Radiative correction to the Casimir energy together with the leading-order Casimir energy for a massive scalar field on the three-edge network, plotted as a function of one edge length (e.g., $L_1$). The remaining edge lengths are fixed at $L_2 = 1$ and $L_3 = 2$. The field mass is $m = 1$ and the coupling constant is $\lambda = 0.1$. Both contributions to the Casimir energy are negative, and the radiative correction is approximately two order of magnitude smaller than the leading-order term.

## IV. RADIATIVE CORRECTION

To obtain the radiative correction to the Casimir energy, we begin with the subtraction of vacuum energies described by the second term in Eq. (14). Substituting the vacuum energy expression in Eq. (13) into Eq. (14) and using the Green's function given in Eq. (A5), we obtain

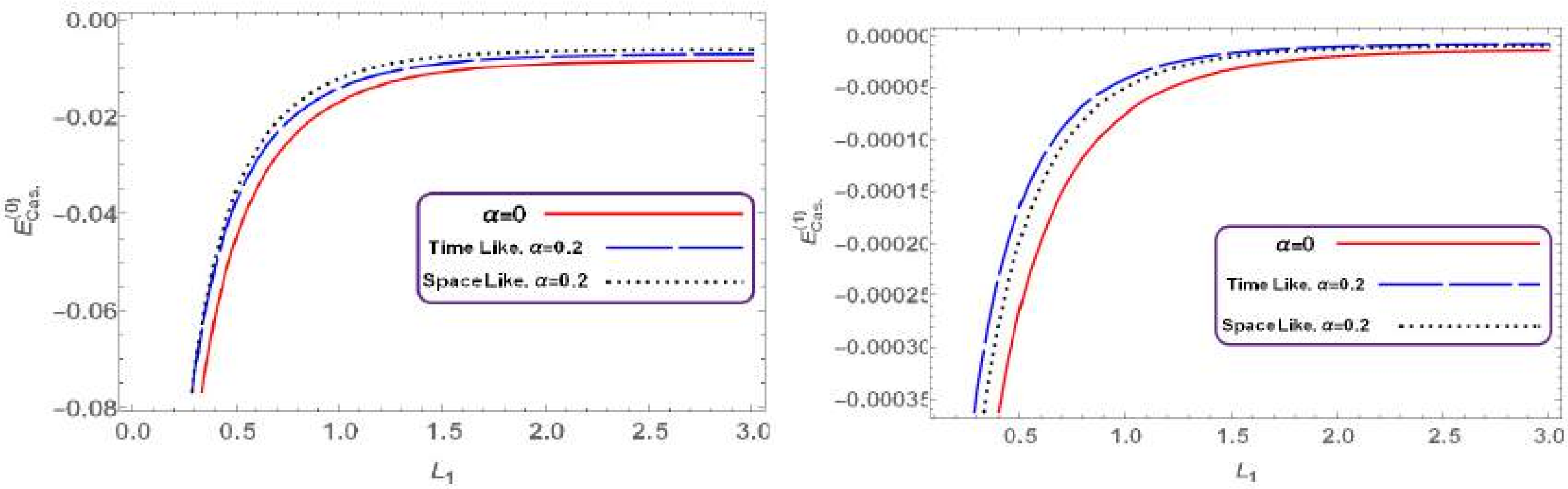

FIG. 4. Left panel: Leading-order Casimir energy for a massive Lorentz-violating scalar field on the three-edge network, plotted as a function of one edge length (e.g., $L_1$). Right panel: First-order radiative correction to the Casimir energy for the same system. In both panels, the remaining edge lengths are fixed at $L_2 = 1$ and $L_3 = 2$, the field mass is $m = 1$, and the coupling constant is $\lambda = 0.1$. The Lorentz-violation parameter is set to $\alpha = 0.2$. Each panel contains three curves: the Lorentz-symmetric case ($\alpha = 0$), time-like Lorentz symmetry breaking, and space-like Lorentz violation. Although Lorentz violation does not change the sign of the Casimir energy, it produces a significant quantitative deviation in both the leading-order and radiative correction terms.

$$
\begin{aligned}
E^{(1)}_{\rm Cas.} &= \sum_{j=1}^{3} E^{(1)}_{{\rm Cas.}j} = \sum_{j=1}^{3}\left[\lim_{L'_j\to\infty}\Delta E^{(1)}_{{\rm vac.}j}\right] \\
&= \sum_{j=1}^{3}\lim_{L'_j\to\infty}\left[\frac{-\lambda}{8}\int_0^{L_j} dx_j G^2(x_j,t;x_j,t) - \{L_j\to L'_j\}\right] \\
&= \sum_{j=1}^{3}\lim_{L'_j\to\infty}\left[\frac{-\lambda}{8}\int_0^{L_j} dx_j \Big(\sum_k f_j(k)\Big)^2 - \{L_j\to L'_j\}\right].
\end{aligned}
\tag{23}
$$

The summation over $k$ in Eq. (23) runs over all roots of the constraint function $\Delta(k) = 0$. Since the roots are neither uniformly spaced nor regularly distributed, direct summation is not feasible. Therefore, we employ Cauchy's theorem to evaluate the sum. This theorem allows us to convert the bracketed expression in Eq. (23) into

$$
\Delta E^{(1)}_{{\rm vac.}j} = \frac{-\lambda}{8}\int_0^{L_j} dx_j\left[\frac{1}{\pi}\int_m^\infty g_j(y)dy + \frac{R}{2\pi}\int_{\frac{-\pi}{2}}^{\frac{\pi}{2}} g_j(Re^{i\theta})e^{i\theta}d\theta + \sum_{n=1}^{\infty}\sum_{i=1}^{3} f_j(p_{n,i})\right]^2 - \{L_j\to L'_j\},
\tag{24}
$$

where $g_j(y) = -f_j(iy)\,\frac{\Delta'(iy)}{\Delta(iy)}$. The second term in the bracket of Eq. (24), which corresponds to the curved part of the semicircular contour in the complex $z$-plane, vanishes in the limit $R\to\infty$. Consequently, only the contribution from the line segment along the imaginary axis (the first integral in the bracket of Eq. (24)) remains. The same computation applies to the auxiliary network. Therefore, the subtraction of vacuum energies in Eq. (24) reduces to

$$
\Delta E^{(1)}_{{\rm vac.}j} = \frac{-\lambda}{8}\int_0^{L_j} dx_j\left[\frac{1}{\pi}\int_m^\infty g_j(y)dy + \sum_{n=1}^{\infty}\sum_{i=1}^{3} f_j(p_{n,i})\right]^2 - \{L_j\to L'_j\}.
\tag{25}
$$

The second term in the bracket of Eq. (25) corresponds to the summation over all poles of the constraint function $\Delta(k)$. This summation is divergent and therefore requires regularization. Since these poles are regularly spaced, we employ the APSF given in Eq. (19) to perform the regularization. Applying the APSF yields

$$
\begin{aligned}
\Delta E^{(1)}_{{\rm vac.}j} = \frac{-\lambda}{8}\int_0^{L_j} dx_j\,&\Bigg\{\frac{1}{\pi}\int_m^\infty g_j(y)dy \\
&+\sum_{i=1}^{3}\Bigg[\underbrace{\frac{-3(x_j-L_j)^2}{4mL_j^3}}_{\mathcal{Z}_{\rm A}} + \underbrace{\frac{2L_i}{\pi}\int_0^\infty \frac{\xi d\xi}{\sqrt{\xi^2+m^2}}\frac{\sin^2[\frac{n\pi}{L_i}(x_j-L_j)]}{2\xi L_j - \sin(2\xi L_j)}}_{\mathcal{I}_{\rm A}} + \mathcal{B}_i(x_j;L_j)\Bigg]\Bigg\}^2 - \{L_j\to L'_j\},
\end{aligned}
\tag{26}
$$

where $\mathcal{Z}_A$ and $\mathcal{I}_A$ denote the zero term and the integral term of the APSF, respectively. These terms, together with their cross term, render Eq. (26) divergent. To remove these divergent contributions, we apply the cutoff regularization scheme, like as the same procedure used in Eq. (20). The subtraction of vacuum energies between the two networks shown in Fig. (1) then ensures that all contributions arising from $\mathcal{Z}_A$, $\mathcal{I}_A$, and their cross term are canceled. Consequently, the only remaining contributions after this regularization and subtraction procedure are the convergent terms:

$$\Delta E^{(1)}_{\text{vac}.j} = \frac{-\lambda}{8}\int_0^{L_j} dx_j \left\{\frac{1}{\pi}\int_m^\infty g_j(y)dy + \sum_{i=1}^{3}\mathcal{B}_i(x_j;L_j)\right\}^2 - \{L_j \to L'_j\}, \tag{27}$$

where $\mathcal{B}_i(x_j;L_j)$ denotes the branch-cut term of the APSF. This term yields a finite contribution and is given by

$$\mathcal{B}_i(x_j;L_j) = \frac{-4L_i}{\pi}\int_m^\infty \frac{\tau}{\sqrt{\tau^2-m^2}}\frac{\sinh^2[\tau(x_j-L_j)]}{2\tau L_j - \sinh(2\tau L_j)}\frac{d\tau}{e^{2\tau L_i}-1}. \tag{28}$$

The final step in the calculation of the radiative correction to the Casimir energy, as indicated in Eq. (23), is to take the limit $L'_j \to \infty$. Applying this limit to Eq. (27) yields the following expression for the first-order radiative correction to the Casimir energy in the three-edge network:

$$E^{(1)}_{\text{Cas.}} = \frac{-\lambda}{8}\sum_{j=1}^{3}\int_0^{L_j} dx_j \left[\frac{1}{\pi}\int_m^\infty g_j(y)dy + \sum_{i=1}^{3}\mathcal{B}_i(x_j;L_j)\right]^2. \tag{29}$$

To obtain the radiative correction to the Casimir energy for a Lorentz-violating scalar field, it is sufficient to set $\alpha \neq 0$ (with $\alpha < 1$) and repeat the computation. This procedure yields the following expression for the radiative correction in the time-like Lorentz-violating case:

$$\tilde{E}^{(1)}_{\text{Cas.Lv}} = \frac{E^{(1)}_{\text{Cas.}}}{1+\alpha}, \tag{30}$$

where $E^{(1)}_{\text{Cas.}}$, given in Eq. (29), corresponds to the Lorentz-symmetric case. For the space-like Lorentz-violating scalar field, the radiative correction to the Casimir energy is obtained as

$$\tilde{E}^{(1)}_{\text{Cas.Lv}}(L_j) = \frac{E^{(1)}_{\text{Cas.}}(\tilde{L}_j)}{\sqrt{1-\alpha}}, \tag{31}$$

where $\tilde{L}_j = \frac{L_j}{\sqrt{1-\alpha}}$. The radiative correction to the Casimir energy is compared with the leading-order term in Fig. (3). Furthermore, the right panel of Fig. (4) compares the radiative correction in the presence and absence of Lorentz violation. This figure shows that, although time-like and space-like Lorentz violation do not change the sign of the correction term, they induce a significant quantitative deviation in the Casimir energy.

## V. CONCLUSION

In this paper, we have computed both the leading-order and the first-order radiative correction to the Casimir energy for a massive scalar field in the $\phi^4$ theory, confined to a simple network consisting of three edges. The scalar field satisfies Dirichlet boundary conditions at the endpoints of each edge. We have also extended our results to the case of a Lorentz-violating scalar field. The main novelty of this work lies in the calculation of the radiative correction to the Casimir energy in the network configuration. A key aspect of our approach is the use of position-dependent counterterms. These counterterms not only renormalize the bare parameters of the Lagrangian but also incorporate the effects of boundary conditions and geometry into the vacuum energy, thereby ensuring a consistent renormalization procedure. To regularize the divergences arising in the calculation of both the leading-order and radiative correction terms, we employed a slightly modified box subtraction scheme. By introducing additional regulators, this method provides a clearer and more transparent way of eliminating infinities. Moreover, supplementing the box subtraction scheme with a cutoff regularization technique further simplifies the removal of divergent contributions. Another important novelty of our work is the consideration of a massive scalar field and the inclusion of Lorentz violation. Previous studies have addressed only the massless case in a Lorentz-symmetric scenario and considered only the leading-order contribution [34]. In contrast, our analysis generalizes these results to the massive and Lorentz-violating cases and includes the first-order radiative correction arising from the self-interaction term in the $\phi^4$ Lagrangian. All obtained results are consistent with physical expectations. In particular, the total Casimir energy remains negative for both the leading-order and the first-order radiative correction terms. Although Lorentz violation does not alter the sign of the Casimir energy, it produces a significant quantitative deviation in both orders, as illustrated in Fig. (4). Future work may extend this analysis to more general network configurations with arbitrary numbers of edges and nodes. It would also be interesting to study other types of boundary conditions, such as Neumann or mixed boundary conditions, and to investigate higher-dimensional network geometries.

## Appendix A: Computation of the Green's Function

In this appendix, we present the details of the computation of the Green's function in the $j$-th edge of the network shown in Fig. (1). To simplify the derivation, we first set the Lorentz-violation parameter to zero ($\alpha = 0$). Starting from Eqs. (2) and (3), we obtain

$$[\partial_t^2 - \partial_{x_j}^2 + m^2]G(x_j, t; x'_j, t') = \delta(t - t')\delta(x_j - x'_j). \tag{A1}$$

The Green's function is initially defined by the expression

$$G(x_j, t; x'_j, t') = \sum_k \int \frac{d\omega}{2\pi} \mathcal{A}_k(x'_j, t')\phi_k(x_j)e^{-i\omega t}, \tag{A2}$$

Here, the summation over $k$ denotes the sum over all roots of the constraint function $\Delta(k) = 0$. By substituting the Green's function in Eq. (A2) into Eq. (A1), multiplying the resulting expression by $\phi^*_{k'}(x'_j)e^{i\omega t'}$, and integrating over $x_j$ from 0 to $L_j$, we obtain the following expression for the coefficient $\mathcal{A}_k$:

$$\mathcal{A}_{k'}(x'_j, t') = \frac{\phi^*_{k'}(x'_j)e^{i\omega t'}}{-\omega^2 + k'^2 + m^2}. \tag{A3}$$

Using Eqs. (A2) and (A3), the Green's function finally takes the form

$$G(x_j, t; x'_j, t') = \sum_k \int \frac{d\omega}{2\pi} \frac{\phi_k(x_j)\phi^*_k(x'_j)e^{-i\omega(t-t')}}{-\omega^2 + k^2 + m^2}. \tag{A4}$$

To substitute the Green's function expression into Eq. (13), we require a simplified form of the Green's function, such as $G(x_j, t; x_j, t)$. Therefore, by performing a Wick rotation and integrating over the frequency $\omega$, the expression in Eq. (A4) reduces to

$$G(x_j, t; x_j, t) = \sum_k f_j(k), \tag{A5}$$

where $f_j(k) = \frac{2k}{\sqrt{k^2+m^2}} \frac{\sin^2[k(x_j - L_j)]}{2kL_j - \sin(2kL_j)}$.

---

* EMail: mavalouyan@iau.ir